# Numerical study on the performance of a glazed photovoltaic thermal system integrated with phase change material (GPVT/PCM): on the contribution of PCM volumetric fraction and environmental temperature

**A Preprint**
June 2021

**Ali Naghdbishi[1], Mohammad Eftekhari Yazdi[2,*], Ghasem Akbari[1]**

[1] *Department of Mechanical Engineering, Qazvin Branch, Islamic Azad University, Qazvin, Iran.*

[2] *Department of Mechanical Engineering, Central Tehran Branch, Islamic Azad University, Tehran, Iran.*

**Abstract**

In the present article, performance of a glazed photovoltaic thermal system integrated with phase change materials (GPVT/PCM) is numerically examined based on both energy and exergy viewpoints. The effect of PCM volumetric fraction and environmental temperature on thermal and electrical characteristics of the system is evaluated. A three-dimensional model of the system is simulated transiently in the ANSYS Fluent 18.2 using pressure-based finite volume method and SIMPLE algorithm selected for coupling pressure and velocity components. Validity of the numerical results is confirmed based on available data. The results indicate that PCM melting is more likely far from the riser tube and absorption of thermal energy be the PCM is mostly effective for filling the PCM container by around 2/3 of the volume. Increase of the PCM volumetric fraction reduces the module temperature and enhances the electrical performance of the system in terms of both energy and exergy efficiencies, while it decreases the thermal efficiency. An opposite trend is experienced by increase of the environmental temperature in which the thermal efficiency enhances and the electrical efficiency declines.

* Corresponding Author: moh.eftekhari_yazdi@iauctb.ac.ir.

**Nomenclature**

| | | | |
|---|---|---|---|
| $A$ | Area (m²) | $\eta$ | Energetic efficiency (%) |
| $C_p$ | Specific heat capacity (J. kg$^{-1}$. K$^{-1}$) | $\rho$ | Density (kg. m$^{-3}$) |
| $d$ | Diameter (m) | $\beta$ | Liquid volumetric fraction |
| $\dot{E}$ | Power (W) | $\varepsilon$ | Exergetic efficiency (%) |
| $\dot{G}$ | Solar irradiation intensity (W. m$^{-2}$) | *Subscripts* | |
| $h$ | Fluid enthalpy (J. kg$^{-1}$) | *amb* | Ambient |
| $H$ | PCM enthalpy (J. kg$^{-1}$) | *C* | Collector |
| $k$ | Thermal conductivity (W. m$^{-1}$. K$^{-1}$) | *el* | Electrical |
| $L$ | Latent heat of PCM (J. kg$^{-1}$) | *g* | Glass cover |
| $\dot{m}$ | Mass flow rate (kg. s$^{-1}$) | *in* | Inlet |
| $P$ | Pressure (kPa) | *out* | Outlet |
| $Q$ | Dissipated heat (W) | *r* | Standard test condition |
| $T$ | Temperature (K) | *s* | Solid |
| $V$ | Velocity (m. s$^{-1}$) | *th* | Thermal |
| *Greeks* | | *w* | Wind |
| $\mu$ | Dynamic viscosity (kg. m$^{-1}$. s$^{-1}$) | | |

# 1 Introduction

Nowadays, by increasing the amount of fossil fuels price and their drawbacks such as global warming, health issues, and environmental problems, countries are more willing to invest in renewable energies. Solar energy is one the most reliable sources which due to its availability, can be used worldwide. Photovoltaic (PV) modules are such a solar system that can directly convert solar energy into electrical energy. However, the electrical performance of the PV modules is dependent on the operating temperature of the PV cells. By increasing the temperature of the PV cells, the electrical output of the module significantly declines [1]. Consequently, cooling of PV panel is an effective approach to enhance its electrical performance.

Water cooling is a conventional method for decreasing the PV cell temperature, that was performed in several configurations, namely submerging the module in water [2], spraying of water over the cell surface [3] or passing an external film of water on the PV surface [4].

Combination of the PV module with a thermal collector, known as photovoltaic thermal (PVT) system, is a more practical method to decrease the cell temperature. This is a compact hybrid system that simultaneously generates both electricity and thermal energy [5]. Numerical study of Bahaidarah et al. [6] indicated that a water-cooled PVT could decrease the cell temperature about 20% leading to 9% enhancement of electrical efficiency compared to the output of a PV panel. Cooling of PVT by organic fluids such as R236fa and R245fa instead of water could reduce the pump power consumption with an electrical efficiency enhancement [7]. Kazemian et al. [8] studied the effect of using glass cover and different working fluids such as water, pure ethylene glycol (EG), and water/EG (50%) on the performance of a PVT system. Based on their experimental research, the water/EG (50%) was the best heat transfer fluid of the PVT system in the cold climates. Besides, the overall energy efficiency of the glazed PVT system was reported to be higher than the unglazed one.

Dispersion of nanoparticles in the heat transfer fluid of a PVT module is an effective method for boosting the system performance. In fact, adding nanoparticles to the base fluid leads to a growth in the thermal conductivity of the mixture which has a favorable effect on the performance of the PVT system [9]. The effect of using three different metal-oxides/water nanofluids on the efficiency of a PVT system was examined by Sardarabadi et al. [10]. According to this study, the PVT system based on the water/$Al_2O_3$ nanofluid has the highest amount of efficiency in comparison with using water/ZnO and water/$TiO_2$ as the heat transfer fluid in the PVT system. Based on three-dimensional simulation, Hosseinzadeh et al. [11] achieved 61.21% relative growth for the thermal efficiency of ZnO/water-cooled PVT panel. Among all the examined parameters in their research, they found coolant inlet temperature as the most effective parameter in enhancement of the thermal efficiency. Numerical and experimental study of Fayaz et al. [12] conducted on a MWCNT/water-cooled PVT module indicated that thermal and electrical efficiencies of this system is up to 5.13% and 12.25% higher than those of a water-cooled PVT panel. Adding various nanoparticles to the base fluid in several other studies also confirmed such performance enhancement, namely dispersion of $Al_2O_3$ and Cu in water and ethylene glycol [13], and adding $Al_2O_3$, CuO, and SiC nanoparticles [14], single-walled carbon nanotubes and graphene nanoplatelets [15], and MWCNT [16] to the water base fluid.

In addition to nanofluids, integration of PVT system with a Phase Change Material (PCM) can improve the system performance [17]. PCM is applicable for thermal regulation by gaining heat from the environment and releasing that energy into its surrounding periodically. Such capability motivated researchers to investigate the effect of using PCM in various fields such as solar energy applications [18-20]. Al-Waeli et al. [21] experimentally studied the performance of a nano-PCM-based PVT system and used SiC nanoparticles in order to increase the conductivity of PCM material. Based on this configuration, electrical efficiency of the system increased from 7.1% to 13.7%. Kazemian et al. [22] conducted an experimental study and reported 4.22% increase of the electrical efficiency after integration of PCM in the EG as the heat transfer fluid. Furthermore, they found that the PV/PCM, water-based PVT/PCM, and EG-based PVT/PCM reduce the amount of entropy generation of the system by 0.58%, 2.42% and 2.84%, respectively, compared to the conventional PV module. Yang et al. [23] also observed that adding PCM to the a simple PVT system increases the thermal efficiency from 58.35% to 69.84% and rises the electrical efficiency from 6.98% to 8.16%. Increase of the melting temperature of PCM was also reported as an assisting parameter in enhancement of the thermal performance of the system [24].

Regarding simultaneous utilization of PCM and nanofluid, Salari et al. [25] investigated the effect of ZnO/water, MWCNT/water, and hybrid nanofluids on the performance of a PVT/PCM system from the energy viewpoint. They found the MWCNT/water-based system as the best configuration with the highest overall efficiency, namely 61.07%. Hosseinzadeh et al. [26] experimentally investigated the effect of using ZnO/water nanofluid as the active coolant fluid and organic paraffin wax as the PCM on a PVT system performance. Enhancement of the output thermal power by approximately 29.60% and the maximum overall exergy efficiency of 13.61% were reported in this research. Numerical simulation of PVT/PCM system integrated with water and $SiO_2$-nanofluid indicated that using PCM can reduce the PVT cell temperature by about 16ºC [27]. A rise of thermal efficiency about 3.51% was also reported for the PVT/PCM system when water/SiO2 nanofluid is used as the coolant fluid. Naghdbishi et al. [28] conducted an experimental study to investigate the effect of MWCNT-water/glycol based nanofluids on a PVT/PCM system. Enhancement of thermal and electrical energetic efficiencies up to 23.58% and 4.21% was reported as the result of dispersion of MWCNT nanoparticles in the water base fluid. A corresponding improvement was also experienced for the electrical and particularly the thermal exergetic efficiencies. Heat dissipation from the PVT panel to surrounding and sun-PVT surface temperature difference were found as the major contributing factors for external exergy losses and exergy destructions of the system, respectively.

According to the literature reviewed, although several numerical and experimental studies have been performed on the investigation of the performance of the PVT/PCM system, there are still some gaps that should be filled. The objective of the present study is to evaluate the effect of environmental temperature and the PCM volumetric fraction on the performance of a GPVT/PCM system from both energy and exergy viewpoints. A three-dimensional model of the system is simulated transiently and validated with the available computational data from the literature. Energetic and exergetic efficiencies are computed in order to evaluate the impact of the target parameters on the system performance in terms of both thermal and electrical attributes.

The article is organized as follows. Problem description and governing equations are introduced in section 2. The energy and exergy analysis for evaluation of the GPVT/PCM system are described in section 3. It follows by mesh independency test and validation of the simulations in section 4. The results are presented and discussed subsequently to evaluate the system performance and its characterizing parameters.

## 2 Numerical modeling

### 2.1 Problem statement

In order to investigate the effect of PCM extent and temporal ambient temperature on the performance of a glazed photovoltaic thermal system integrated with phase change material (GPVT/PCM), a three-dimensional model of the system is simulated. The purpose is to evaluate thermal and electrical characteristics of the system regarding energy and exergy analyses.

The system involves a glass cover, two layers of EVA, a photovoltaic module, a Tedlar layer, and an absorber layer. Furthermore, riser tubes and a PCM container are attached beneath the GPVT system, in order to cool the PV module and provide thermal energy. The actual collector involves an array of parallel tubes, but in order to reduce the computational cost of 3D simulations, just one tube with its neighbor region is considered such that its side walls coincide with the symmetry mid-plane located between each pair of riser tubes. A schematic diagram of the designed GPVT/PCM system confined to the mentioned computational domain is illustrated in Figure 1(a). The dimensions of each layer and their thermodynamic properties are presented in Table 1. It should be noted that the anti-reflection coating layer has a tremendously small thickness (0.0001 mm), which is neglected in the simulation. It is assumed that the thermal properties of all materials are constant and do not change with temperature. The properties and dimensions of the PCM and riser tube are also presented in Table 1. In order to investigate the effect of PCM extent on the performance of the PVT/PCM system, five equidistant dividers are considered creating six identical partitions. These dividers are demonstrated in Figure 1(b).

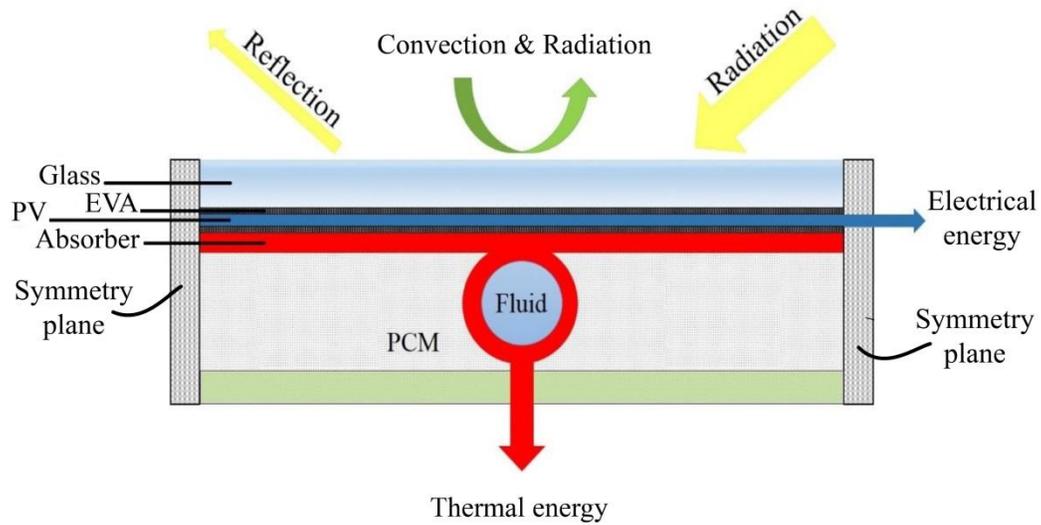

**(a)**

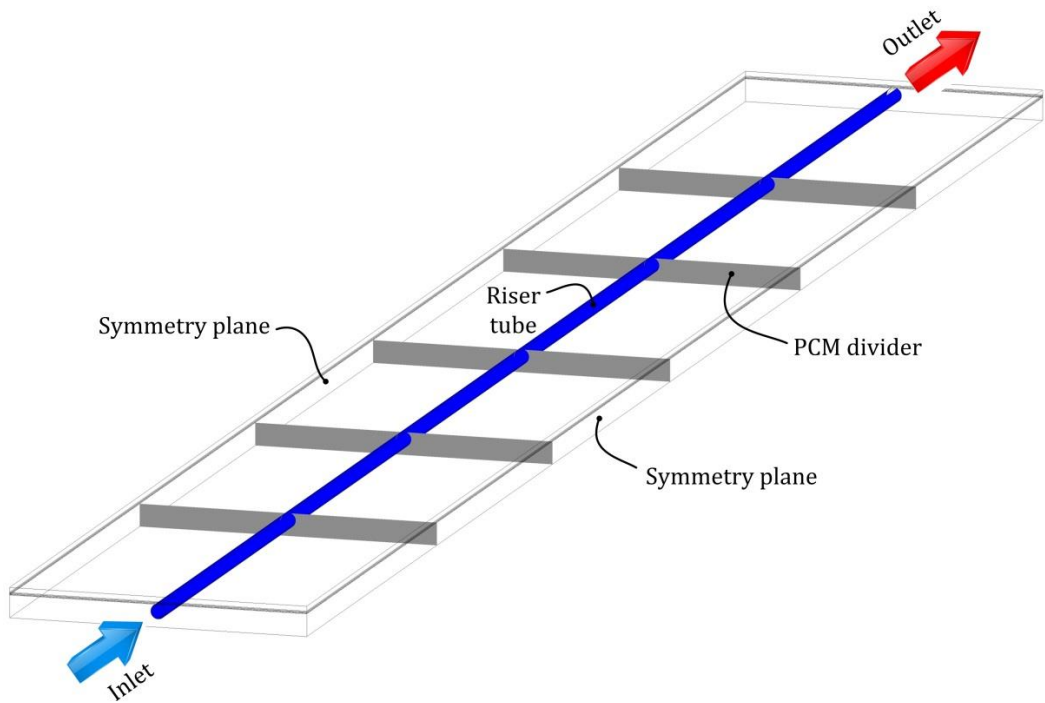

**(b)**

**Figure 1**. (a) Schematic illustration of the designed GPVT/PCM system and the computational domain considered; (b) Geometry of the PCM container including five dividers to create six identical partitions.

**Table 1.** Thermodynamic properties and dimensions of the simulated GPVT/PCM system components [29, 30].

| Element | Density (kg. m$^{-3}$) | Thermal conductivity (W. m$^{-1}$. K$^{-1}$) | Specific heat capacity (J. kg$^{-1}$. K$^{-1}$) | Dimensions (cm) | Melting temperature (°C) / Enthalpy of fusion (kJ. kg$^{-1}$) |
|---|---|---|---|---|---|
| Glass cover | 2200 | 0.76 | 830 | 164×20×0.32 | |
| EVA | 960 | 0.35 | 2090 | 164×20×0.05 | |
| PV cells | 2330 | 148 | 700 | 164×20×0.03 | |
| Tedlar | 1200 | 0.2 | 1250 | 164×20×0.01 | |
| Absorber | 8960 | 401 | 385 | 164×20×0.04 | |
| PCM | 800 | 0.25 | 2300 | 164×20×1.5 | 55 / 170 |
| Riser tube | 8960 | 401 | 385 | Inner diameter: 0.8 Outer diameter: 1 | |

## 2.2 Governing equations

The mass, momentum and energy equations for the heat transfer fluid is expressed as:

$$\frac{\partial \rho_f}{\partial t} + \nabla \cdot (\rho_f \boldsymbol{V}_f) = 0 \tag{1}$$

$$\rho_f \left[ \frac{\partial \boldsymbol{V}_f}{\partial t} + (\boldsymbol{V}_f \cdot \nabla)\boldsymbol{V}_f \right] = -\nabla P + \nabla \cdot (\mu_f \nabla \boldsymbol{V}_f) \tag{1}$$

$$\rho_f C_{p,f} \frac{\partial T_f}{\partial t} + \rho_f C_{p,f} \boldsymbol{V}_f \cdot \nabla T_f = \nabla \cdot (k_f \nabla T_f) \tag{3}$$

where subscript $f$ denotes the heat transfer fluid and $P$, $\boldsymbol{V}$, $\rho$ and $\mu$ are fluid pressure, velocity vector, density and dynamic viscosity, respectively. The energy equation for the solid domains are:

$$\rho_s C_{p,s} \frac{\partial T_s}{\partial t} = \nabla \cdot (k_s \nabla T_s) \tag{4}$$

where $s$ indicates the solid regions. Regarding the PCM domain, the mass and momentum equations are as follows [31]:

$$\frac{\partial \rho_{PCM}}{\partial t} + \nabla \cdot (\rho_{PCM} \cdot \boldsymbol{V}_{PCM}) = 0 \tag{5}$$

$$\rho_{PCM} \left[ \frac{\partial \boldsymbol{V}_{PCM}}{\partial t} + (\boldsymbol{V}_{PCM} \cdot \nabla)\boldsymbol{V}_{PCM} \right] = -\nabla P + \nabla \cdot (\mu_{PCM} \nabla \boldsymbol{V}_{PCM}) + \boldsymbol{S} \tag{6}$$

The source term $\boldsymbol{S}$ is defined as:

$$\boldsymbol{S} = \frac{(1-\beta)^2}{(\beta^2 + \phi)} A_{mush}(\boldsymbol{V}_{PCM} - \boldsymbol{V}_P) \tag{7}$$

where $\beta$, $A_{mush}$, and $\boldsymbol{V}_p$ are the liquid volume fraction, mushy zone constant (considered to be $10^5$), and the solid velocity, respectively. $\phi$ is a very small number that is set to 0.001 [31]. The energy equation of the PCM is given as:

$$\frac{\partial}{\partial t}(\rho_{PCM} H_{PCM}) + \nabla \cdot (\rho_{PCM} \boldsymbol{V}_{PCM} H_{PCM}) = \nabla \cdot (k_{PCM} \nabla T_{PCM}) + S \tag{8}$$

where $H$ is enthalpy of the PCM expressed as:

$$H = h_{ref} + \int_{T_{ref}}^{T} C_{p,PCM} dT + \beta L \tag{9}$$

The governing equations are solved by ANSYS Fluent software 18.2. Regarding the boundary conditions, prescribed mass flow (with uniform velocity profile) and temperature are assigned at the inlet of the riser tube. At the tube outlet, "pressure outlet" boundary condition is utilized and no-slip condition is implemented at the inner surface of the tube. Symmetry boundary condition is set for the side walls and adiabatic boundary condition is considered at the bottom surface of the PCM container. Over the surface of the glass cover, both convective and radiative heat transfers are involved. The convective heat transfer coefficient ($h$) is modeled as follows [32]:

$$h = 5.7 + 3.8 V_w \quad \text{if} \quad V_w < 5 \frac{m}{s} \tag{10}$$

where $V_w$ is the wind speed, which is considered to be 1 m/s. The effect of solar irradiation absorbed by the PV cells is modelled as a heat generation in the PV panel. The sky is considered as a black body with a temperature ($T_{sky}$) expressed as [33]:

$$T_{sky} = 0.0522 \cdot T_{amb}^{1.5} \tag{11}$$

where $T_{amb}$ denotes the ambient temperature. The system is simulated unsteadily by setting the environmental temperature as the initial condition.

## 3 Evaluation of the system performance

### 3.1 Energy analysis

The energy balance for the GPVT/PCM system is expressed as follows:

$$\dot{E}_s = \dot{E}_{el} + \dot{E}_{th} + \dot{E}_{loss} \tag{12}$$

where $\dot{E}_{el}$ and $\dot{E}_{th}$ are the electrical power and thermal output of system, respectively. Furthermore, $\dot{E}_{loss}$ indicates the rate of heat losses from the PVT surface to the ambient and $\dot{E}_{sun}$ represents the absorbed solar energy subjected to irradiation $\dot{G}$, modeled as [11]:

$$\dot{E}_s = \alpha \tau A \dot{G} \tag{13}$$

where $\alpha$, $\tau$ and $A$ are the glass cover transmissivity, photovoltaic cells absorptivity and the surface area, respectively. The thermal energy output is characterized by the heat dissipated to the coolant fluid ($\dot{E}_f$) and the thermal energy stored by the PCM ($\dot{E}_{PCM}$):

$$\dot{E}_{th} = \dot{E}_f + \dot{E}_{PCM} \tag{14}$$

where $\dot{E}_f$ and $\dot{E}_{PCM}$ are expressed as follows:

$$\dot{E}_f = \dot{m}_f C_{p,f}(T_{out} - T_{in}) \tag{15}$$

$$\dot{E}_{PCM} = \frac{\dot{m}_{PCM} C_{p,PCM}(T_{MS} - T_0)}{t_{MS}} + \frac{\dot{m}_f C_{p,f}(T_{ME} - T_{MS})}{t_{ME} - t_{MS}} + \frac{\dot{m}_{PCM} C_{p,PCM}(T_{SE} - T_{ME})}{t_{SE} - t_{ME}} \tag{16}$$

where $T_{out}$, $T_{in}$ and $T_0$ are inlet, outlet and initial temperature of the heat transfer fluid, respectively. Additionally, $t_{MS}$, $t_{ME}$, and $t_{SE}$, are the times corresponding to start of the melting process, end of the melting process and end of the simulation, respectively.

Electrical and thermal efficiencies of the system can be calculated as follows [34]:

$$\eta_{el} = \frac{\dot{E}_{el}}{\dot{E}_s} = \eta_r[1 - 0.0045(T_{cell} - 298.15)] \tag{17}$$

$$\eta_{th} = \frac{\dot{E}_{th}}{\dot{E}_s} = \frac{\dot{E}_f + \dot{E}_{PCM}}{\alpha \tau A \dot{G}} \tag{18}$$

Equation (17) is an empirical correlation presented by Evans for the electrical efficiency [34]. $T_{cell}$ and $\eta_r$ are the PV cell temperature and PV module efficiency at the standard test condition, respectively. In the present study, $\eta_r$ is set to be 15% based on the range of commercial PV modules [35].

## 3.2 Exergy Analysis

The exergy balance for the GPVT/PCM system can be stated as follows:

$$\dot{E}x_s = \dot{E}x_{el} + \dot{E}x_{th} + \dot{E}x_{loss} \tag{19}$$

where $\dot{E}x_s$, $\dot{E}x_{el}$, $\dot{E}x_{th}$ and $\dot{E}x_{loss}$ are the sun input exergy to the system, electrical exergy output, thermal exergy output, and exergy loss of the GPVT/PCM system, respectively. The input solar exergy is calculated by [36]:

$$\dot{E}x_s = \dot{E}_s \cdot \left(1 - \frac{T_{amb}}{T_{sun}}\right) \tag{20}$$

where $T_{sun}$ is the sun temperature (considered to be 5800 K), and the electrical and thermal exergy outputs of the system are expressed as:

$$\dot{E}x_{el} = \dot{E}_{el} \tag{21}$$

$$\dot{E}x_{th} = \dot{m}_f[C_{p,f}(T_{out} - T_{in}) - T_{amb}(s_{out} - s_{in})] \tag{22}$$

For the heat transfer fluid, the difference between inlet and outlet entropies is given as:

$$s_{out} - s_{in} = C_{p,f} \ln\left(\frac{T_{out}}{T_{in}}\right) \tag{23}$$

The electrical and thermal efficiencies of the GPVT/PCM system is also calculable by dividing their corresponding exergies to the input solar exergy:

$$\varepsilon_{el} = \frac{\dot{E}x_{el}}{\dot{E}x_s} \tag{24}$$

$$\varepsilon_{th} = \frac{\dot{E}x_{th}}{\dot{E}x_s} \tag{25}$$

## 4 Mesh independency and validation

Increase of computational grid resolution enhances the simulation accuracy but increases computational cost as well. A mesh independency analysis is required to find an optimal configuration regarding the number of computational cells. Six different grids are applied to the GPVT/PCM system, including 300, 700, 1400, 2800, 4000, and 5200 thousand cells. Table 2 presents the simulation results obtained for the outlet temperature of the riser tube and the amount of melted PCM to evaluate the effect of grid resolution on the accuracy. The relative error between each value and that of the finest grid, i.e. 5200 thousand cells, is also given in the table. Based on the results, GPVT/PCM system with 4000 thousand cells reveals a reasonable accuracy with relative error of 0.02% and 0.61% for the outlet temperature and melted PCM, respectively. This configuration is consequently applied for the rest of our simulations.

In order to ensure validity of the obtained results, a numerical simulation based on the present methodology is performed and compared by the numerical study of Su et al. [37] on a PVT/PCM system. For a meaningful comparison, the same thermophysical and operating conditions as well as identical dimensions are considered for both simulations. It consists of a $1.2m \times 0.5m$ panel equipped with a PCM by $6.5\ cm$ thickness, which its enthalpy of fusion and melting temperature are $210\ kJ/kg$ and $80\ °C$, respectively. The panel surface temperature based on our simulations is compared in Figure 2 with the computed temperature reported by [37]. As observed, a good agreement is obtained between simulated results of both studies during the entire time interval. In order to compare the results more quantitatively, the root mean squared error normalized by the measured surface temperature is calculated as follows:

$$E_{RMS} = \left[\frac{1}{N}\left(\sum_{i=1}^{N}\left(\frac{T_{s,i}^{pr} - T_{s,i}^{ref}}{T_{s,i}^{ref}}\right)^2\right)\right]^{1/2} \tag{26}$$

where the $pr$ and $ref$ superscripts denote simulated data by the present study and numerical data presented by Su et al. [37], respectively and $N$ is the number of samples. $E_{RMS}$ based on equation (26) is calculated to be 0.0228 or 2.28%, that indicates reasonable accuracy and reliability of the simulations.

**Table 2.** Evaluation of mesh independency based on six grid configurations.

| Number of cells | Outlet temperature | | Melted PCM | |
|---|---|---|---|---|
| | Value (°C) | Relative error (%) | Value (Kg) | Relative error (%) |
| 300,000 | 35.40 | 0.59 | 1.29 | 20.37 |
| 700,000 | 35.47 | 0.39 | 1.41 | 12.96 |
| 1,400,000 | 35.55 | 0.16 | 1.53 | 5.55 |
| 2,800,000 | 35.59 | 0.05 | 1.60 | 1.23 |
| 4,000,000 | 35.60 | 0.02 | 1.61 | 0.61 |
| 5,200,000 | 35.61 | | 1.62 | |

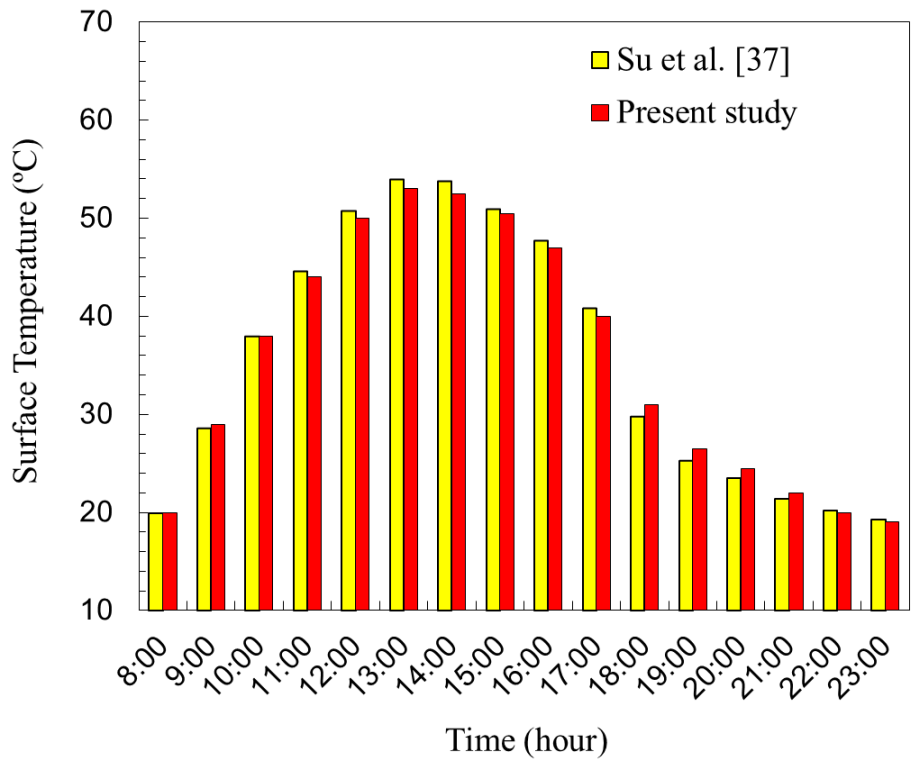

**Figure 2.** Comparison of the simulated surface temperature based on the present methodology and that of the numerical study of Su et al. [37] for a PVT/PCM system.

## 5 Results

### 5.1 GPVT/PCM parameters

Figure 3 illustrates variation of the outlet temperature of heat transfer fluid with time for six different extent of PCM. The fractions notified in this and the subsequent figures specify the volumetric fraction of the container capacity filled by PCM. As it is expected, the fluid temperature at the riser outlet increases gradually, that is due to growth of module temperature by time. In contrast, the fluid temperature decreases by filling the container with more extent of PCM. In fact, by increasing the amount of PCM used beneath the PV module, higher thermal energy is absorbed by the PCM, leading to lower heat dissipated to the working fluid and reduction of its temperature. The same trend is observed in Figure 4 that demonstrates temporal variation of the panel surface temperature for different amount of PCM. Increase of PCM volumetric fraction from 1/6 to 6/6 causes around 2 ºC drops in the surface temperature from 45.29 ºC to 43.33 ºC at the end of simulation time.

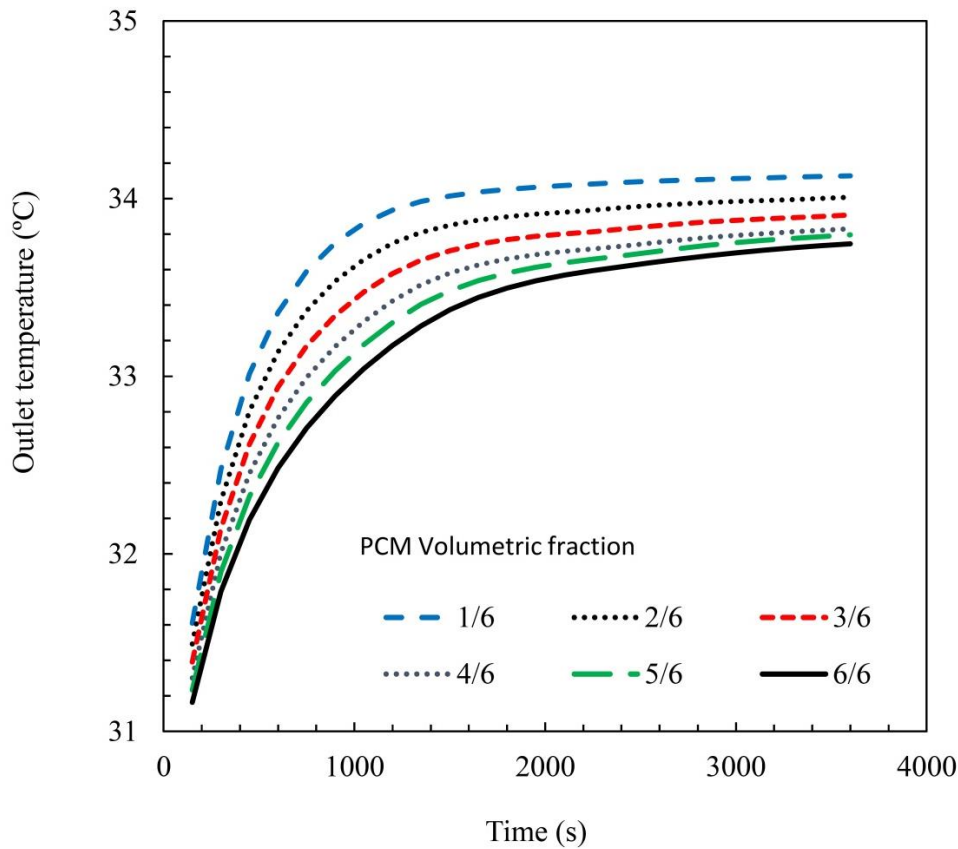

**Figure 3.** Temporal variation of the outlet temperature of heat transfer fluid for different PCM volumetric fractions.

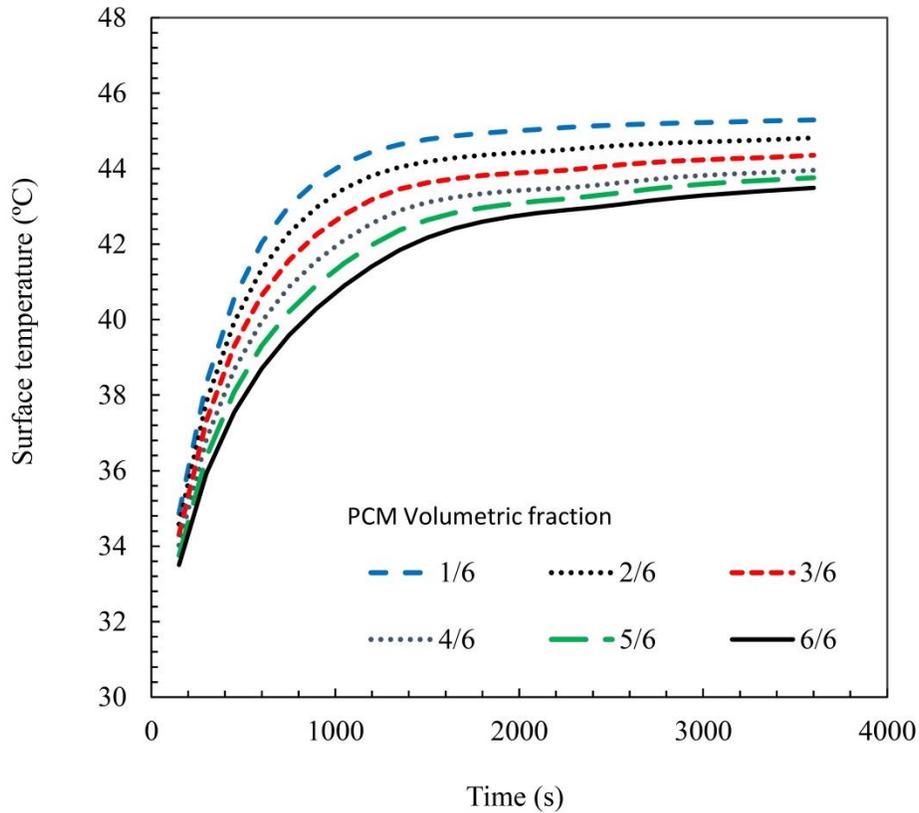

**Figure 4.** Temporal variation of the surface temperature of PVT/PCM system for different PCM volumetric fractions.

Our results indicated that at the initial stage of transient simulation, the PCM remains at the solid state. This is the period in which the PCM temperature steadily increases by sensible heat transfer mechanism. In order to investigate the PCM melting evolution, Figure 5 illustrates the contour-plot of mass fraction of the melted PCM in the lower surface of the GPVT/PCM system. The container is completely filled by PCM (volumetric fraction of 6/6) and four different simulation times, namely 15, 30, 45 and 60 minutes, are considered. The panel orientation and the fluid flow direction are the same as that depicted in Figure 1(b). It is obvious that the PCM melting is originated at the corners located far from the cooling riser tube and closer to its outlet. This is the location that is expected to have the highest cell temperature due to lower capability of the outlet coolant fluid in absorption of thermal energy. Comparison of different contour-plots in Figure 5 shows that the liquid-solid interface of PCM gradually moves toward the riser tube by passing time. However, even after reaching the steady state condition (passing about 60 minutes), the melted PCM does not penetrate throughout the whole content of PCM. In fact, the region at the proximity of the riser tubes is dominantly characterized by the heat dissipated to the coolant fluid and consequently, the PCM temperature could not reach to the melting point at such regions. Furthermore, effective absorption of thermal energy by the coolant fluid at the near-inlet regions causes domination of non-melted PCM at that region.

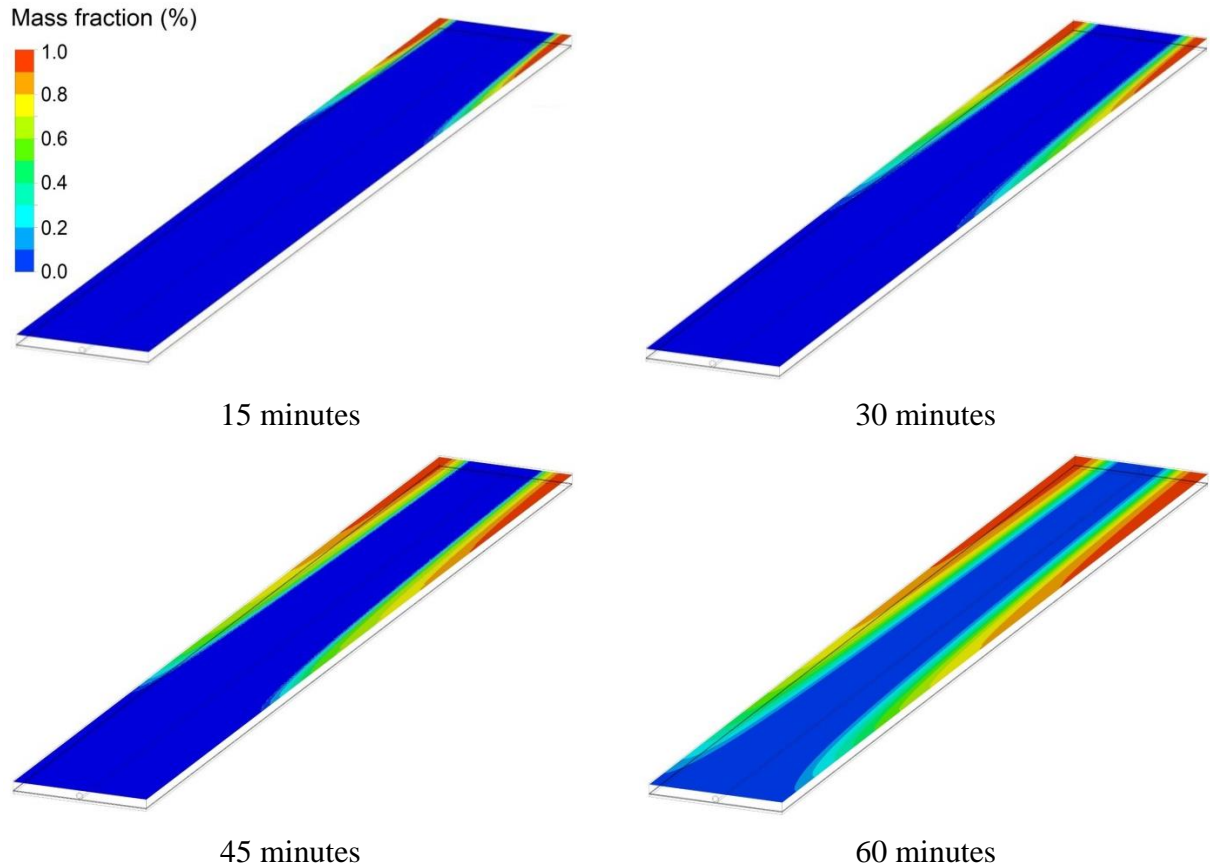

15 minutes     30 minutes

45 minutes     60 minutes

**Figure 5.** Mass fraction of the melted PCM in the lower surface of the GPVT/PCM system at different simulation times.

Figure 6 demonstrates the mass of melted PCM in the steady-state condition for six different PCM volumetric fractions. The melted mass increases effectively from 0.10 Kg to 0.24 Kg, by changing the PCM volumetric fractions from 1/6 to 4/6, whereas it just has a minimal increase to 0.26 Kg by filling the remaining 2/6 volumetric capacity of the PCM container.

In order to evaluate the impact of partial filling of the container by PCM on the temperature of PVT/PCM system, Figure 7 demonstrates contour-plot of temperature in the lower surface of a GPVT/PCM system for two different configurations: (a) with volumetric fraction of 3/6 by filling the three upper partitions of the container (which are nearer to the tube outlet); (b) complete filling of the container by PCM. For the first configuration, the surface temperature steadily increases from the inlet up to the middle section of the panel. From this point forward, a considerable drop in the cell temperature occurs which is due to absorption of thermal energy by the PCM. In contrast, the second configuration with volumetric fraction of 6/6 involves a complete control of the temperature at entire length of the panel. Gradual rise of cell temperature alongside with the riser tube is due to increase of the heat transfer fluid temperature. It is also noticeable in both configurations that the centerline of the panel and its proximity are dominantly attributed by cooling effect imposed by the riser tube.

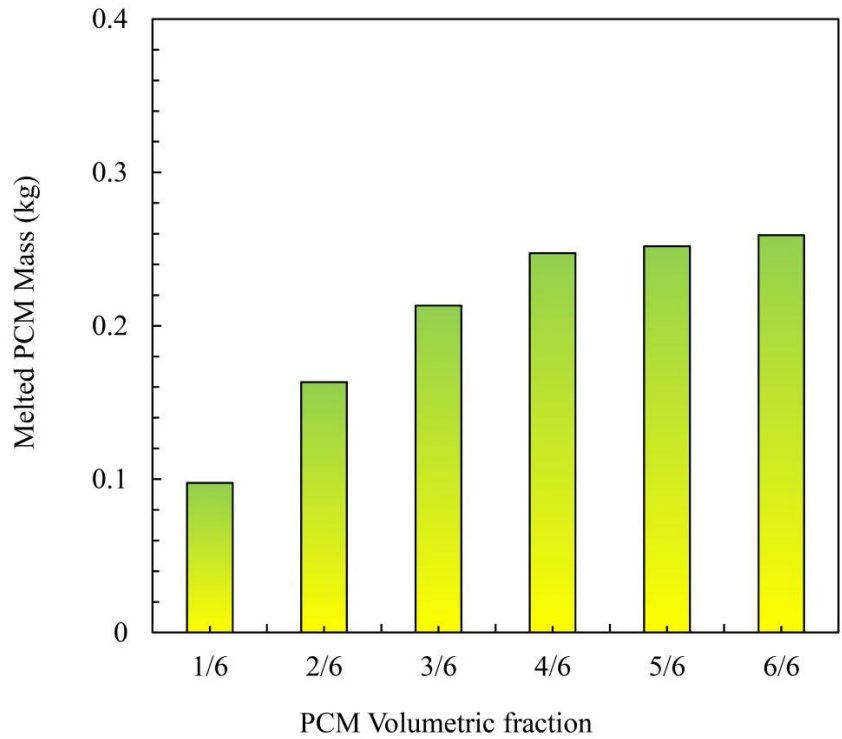

**Figure 6.** Melted PCM mass of GPVT/PCM system at the steady state condition for different volumetric fractions.

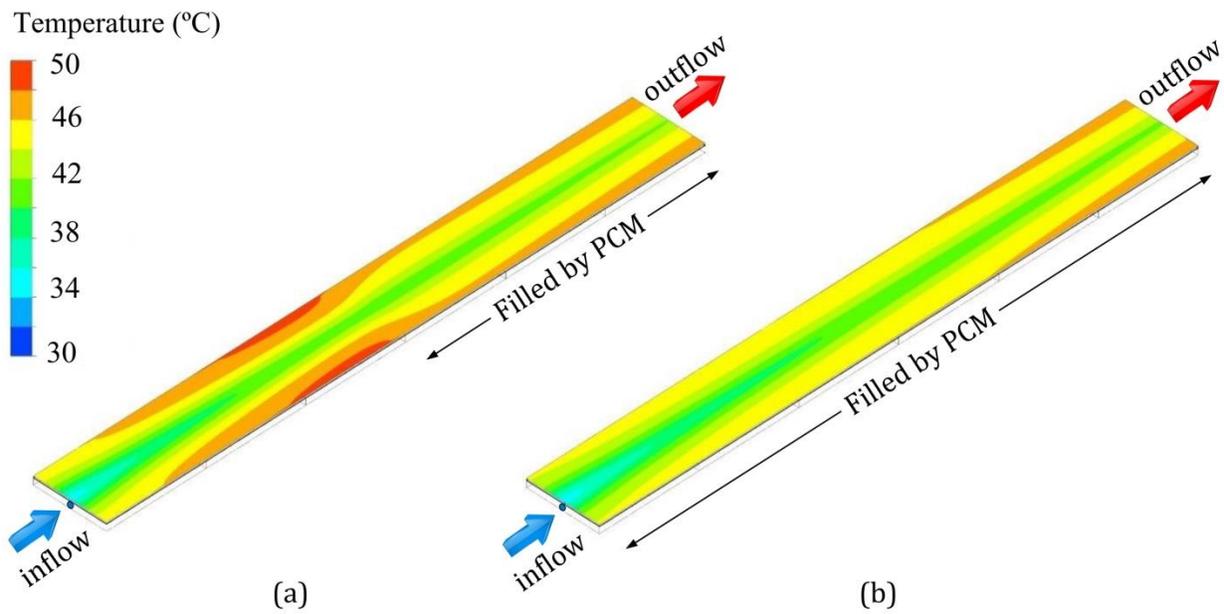

**Figure 7.** Temperature distribution in the lower surface of the GPVT/PCM system for two different volumetric fractions: (a) 3/6 (b) 6/6.

The impact of ambient temperature on the GPVT/PCM system is also investigated. Figure 8 presents temporal variation of the panel surface temperature for three different ambient temperatures, namely 25ºC, 30ºC and 35ºC. For all cases, temperature continually rises to reach the steady state condition. By increase of the ambient temperature from 25 ºC to 35 ºC, the steady state surface temperature of the GPVT/PCM system elevated from 42.77 ºC to 45.82 ºC. In fact, rising the ambient temperature decreases the difference between the surface temperature of the system and the ambient temperature and consequently reduces the heat transfer potential towards the surrounding leading to increase the system temperature. Regarding the outlet temperature of the coolant fluid, the same trend occurs, which is not illustrated here. More specifically, a 10 ºC increase of the environmental temperature from 25 ºC to 35 ºC, rises the outlet temperature of the working fluid from 33.60 °C to 34.19 °C.

Regarding the impact of ambient temperature on the content of melted PCM, our transient simulation indicated that melting originates at times 900 s, 1200 s and 1700 s for the environmental temperatures 35 °C, 30 °C, and 25 °C, respectively. Earlier melting of PCM for higher ambient temperatures is because of less effective heat loss from the PV system in such cases, that rises the system temperature and enhances the energy transferred to the PCM. Table 3 presents the effect of ambient temperature on the mass of the melted PCM at the steady state condition. A continual increase of the melted mass from 0.11 Kg to 0.31 Kg is obtained by rising the environmental temperature from 25 °C to 35 °C.

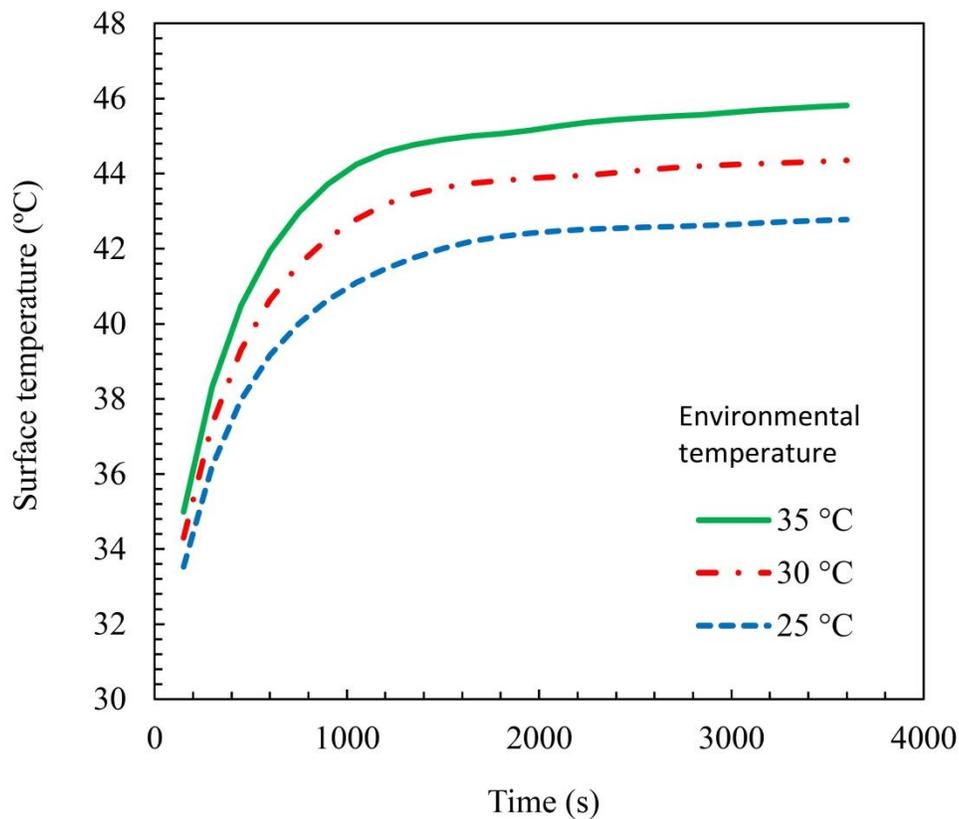

**Figure 8.** Temporal variation of the surface temperature of GPVT/PCM system for different ambient temperatures.

**Table 3.** Melted PCM mass of the GPVT/PCM system at the steady state condition for different ambient temperatures.

| Ambient temperature (°C) | Melted mass (Kg) |
|---|---|
| 25 | 0.11 |
| 30 | 0.21 |
| 35 | 0.31 |

### 5.2 Energetic/exergetic efficiency of GPVT/PCM system

Figure 9 illustrates thermal and electrical energetic efficiencies of the GPVT/PCM system for six different PCM volumetric fractions. It is obvious that by utilizing more extent of PCM, the thermal efficiency of the system reduces whereas the electrical efficiency enhances. Although using more PCM increases the amount of heat absorbed by the PCM from the PV system, it reduces the extent of thermal energy dissipated to the coolant fluid. The overall effect is reduction of sum of these two thermal energies, as the numerator in the definition of the thermal energy efficiency (equation 18), that causes decrease of the thermal efficiency. In contrast, the PV cell temperature reversely contributes to the electrical efficiency (equation 17). Consequently, the cell temperature drop caused by utilizing a higher PCM volumetric fraction results in enhancement of electrical energy efficiency.

Regarding the effect of PCM content on the exergetic efficiency, Figure 10 illustrates thermal and electrical exergetic efficiencies for six different volumetric fractions. Similar to the trends observed for energetic efficiencies, by increase of the PCM extent, the electrical exergy efficiency of the GPVT/PCM system enhances while the thermal exergy efficiency reduces. More specifically, an increase of the electrical exergetic efficiency from 14.42% to 14.55% is observed by elevation of the PCM volumetric fraction from 1/6 to 6/6, while the respective decrease of the thermal exergetic efficiency is from 0.41% to 0.34%.

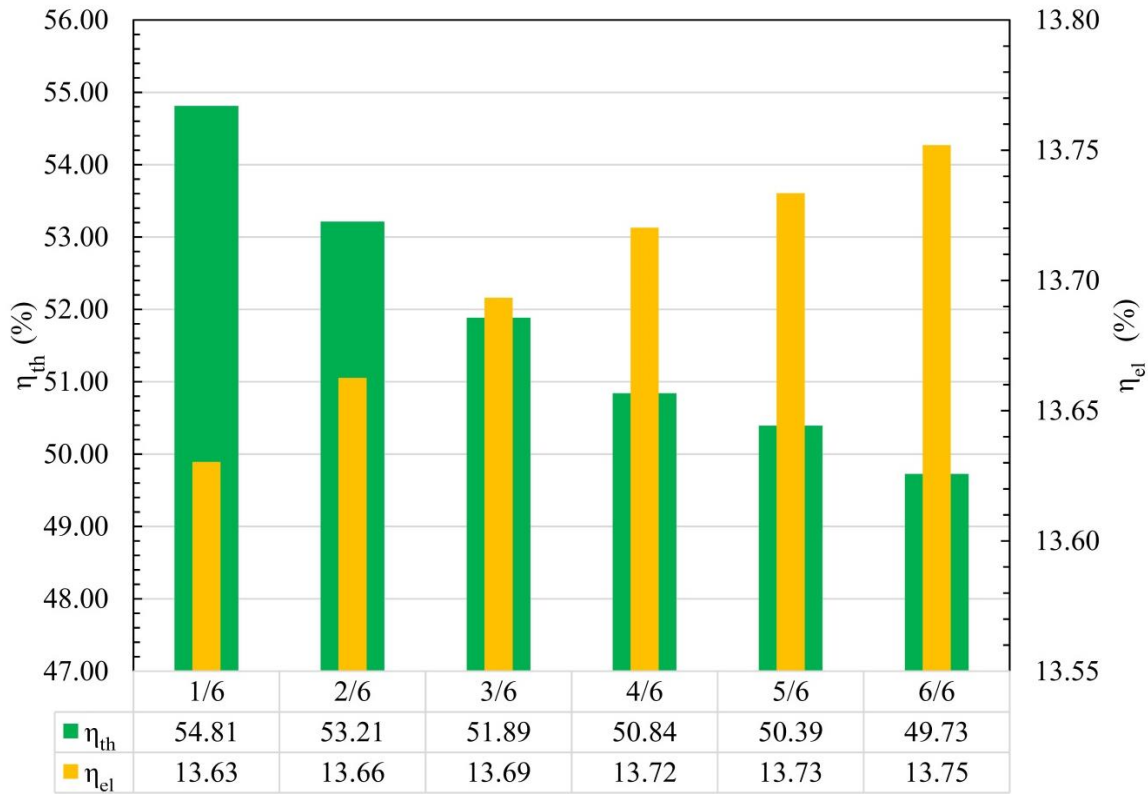

**Figure 9.** The thermal and electrical energy efficiency of the GPVT/PCM system for different PCM volumetric fractions.

The thermal and electrical efficiency of the GPVT/PCM system at various ambient temperatures is presented in Table 4 for both energetic and exergetic viewpoints. The results indicate that 10 ºC rise of environmental temperature increases thermal energetic efficiency from 47.85% to 55.57% and also enhances thermal exergetic efficiency. The increase of temperature of heat transfer fluid is responsible for such elevations. In contrast, both electrical energetic and exergetic efficiencies of the system are reduced by ambient temperature rise due to the rise of GPVT/PCM system temperature. This decrease is from 13.80% to 13.59% in terms of the electrical energetic efficiency, and from 14.60% to 14.39% for the electrical exergetic efficiency.

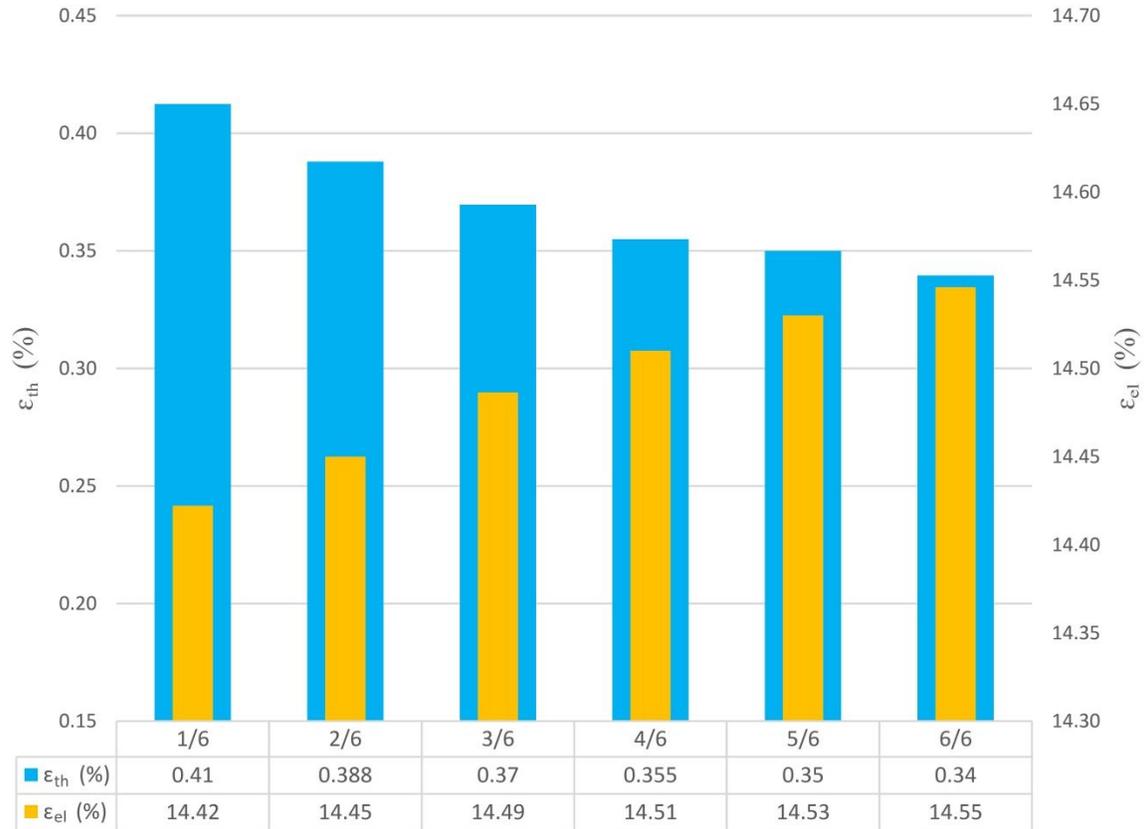

**Figure 10.** The thermal and electrical exergy efficiency of the GPVT/PCM system for different PCM volumetric fractions.

**Table 4.** Thermal and exergetic efficiencies of the GPVT/PCM system for different ambient temperatures from energy and exergy viewpoints.

| Ambient temperature (°C) | Energetic efficiency (%) | | Exergetic efficiency (%) | |
|---|---|---|---|---|
| | Thermal efficiency | Electrical efficiency | Thermal efficiency | Electrical efficiency |
| 25 | 47.85 | 13.80 | 0.31 | 14.60 |
| 30 | 51.89 | 13.69 | 0.37 | 14.49 |
| 35 | 55.57 | 13.59 | 0.42 | 14.39 |

# 6 Conclusion

Three-dimensional transient simulation of a GPVT/PCM system was performed in order to investigate the effect of PCM extent and the ambient temperature on the system performance, both in energy and exergy viewpoints. Validity of the numerical results was confirmed by comparison of the simulated temperature of a PVT/PCM system with the numerical results provided by Su et al. [37]. The most remarkable findings of the present study can be summarized as follows:

- Using a higher volumetric fraction of PCM increased the heat storage capacity of the system and reduced both the cell temperature and the outlet temperature of the coolant fluid. More specifically, increase of PCM volumetric fraction from 1/6 to 6/6 decreased the surface temperature by approximately 2 ºC. Furthermore, in the case of partial filling of the container by PCM, a considerable temperature drop was observed at the proximity of the PCM-filled partitions.

- Effective melting of PCM was experienced by a range of volumetric fraction between 1/6 to 4/6, while adding more PCM marginally increased the mass of melted PCM. Melting process originated at locations far from the cooling riser tube and closer to its outlet. It was also observed that the region at the neighborhood of the riser tubes is dominantly affected by the heat dissipated to the heat transfer fluid, which prevents melting of PCM at that domain at the steady state condition.

- Rise of environmental temperature increased the steady state temperature of the module surface as well as the outlet flow temperature as the consequence of decrease in the heat transfer potential towards the surrounding. It also increased the extent of melted PCM.

- Based on both energy and exergy viewpoints, using a higher volumetric fraction of PCM beneath the PVT system enhanced the electrical efficiency of the system due to reduction of the module temperature. This however decreased the thermal energetic and exergetic efficiencies. An opposite trend was observed for the contribution of ambient temperature. Both energy and exergy analyses indicated that thermal efficiency enhances by increase of environmental temperature while the electrical efficiency of the system deteriorates by such environmental change.